\documentclass[twocolumn,showpacs,prl,amsmath,amssymb]{revtex4}

\usepackage{graphicx}% Include figure files
\usepackage{dcolumn}% Align table columns on decimal point
\usepackage{bm}% bold math

\begin{document}

\title{A model of Hall reconnection}

\author{Leonid M. Malyshkin}
\email{leonmal@uchicago.edu}
\affiliation{Department of Astronomy \& Astrophysics,
University of Chicago, 5640 S. Ellis Ave., Chicago, IL 60637}
\date{\today}

\begin{abstract}
The rate of quasi-stationary, two-dimensional magnetic reconnection is 
calculated in the framework of incompressible Hall magnetohydrodynamics (MHD),
which includes the Hall and electron pressure terms in the Ohm's law.
The Hall-MHD equations are solved in a local region across the reconnection 
electron layer, including only the upstream region and the layer center. In 
the case when the ion inertial length $d_i$ is larger than the Sweet-Parker 
reconnection layer thickness, the dimensionless reconnection rate is found 
to be independent of the electrical resistivity and equal to $d_i/L$, where 
$L$ is the scale length of the external magnetic field in the upstream region 
outside the electron layer, and the ion layer thickness is found to be $d_i$.
\end{abstract}

%52.25.Xz 	Magnetized plasmas (Plasma properties)
%52.30.Ex	Two-fluid and multi-fluid plasmas (Plasma dynamics and flow)
%52.35.Vd 	Magnetic reconnection (Waves, oscillations, and instabilities in plasmas and intense beams)
%94.30.cp 	Magnetic reconnection (Physics of the magnetosphere)
%96.60.Iv 	Magnetic reconnection (Solar physics)

\maketitle

%---------------------------------------------------------------------

\section{\label{INTRODUCTION}
Introduction
}
Magnetic reconnection is a fundamental process of breaking and 
topological rearrangement of magnetic field lines in magnetized 
plasmas. Reconnection converts magnetic energy into kinetic and 
thermal energy and is believed to be responsible for many phenomena 
observed in the laboratory and cosmic 
plasmas~\citep{biskamp_2000,kulsrud_2005}. Because electrical 
resistivity is very low in hot plasmas, magnetic reconnection 
due to resistive dissipation of magnetic field is typically a very 
slow process~\citep{biskamp_2000,kulsrud_2005,yamada_2009}. 
Reconnection can become much faster in the case when 
resistivity is anomalously high due to local plasma 
instabilities~\citep{kulsrud_2001,malyshkin_2005,yamada_2009}. 
Another possibility is fast reconnection made possible because of 
two-fluid plasma effects~\citep{biskamp_2000}, which require a 
two-fluid magnetohydrodynamics (MHD) description of plasma. In the 
limit of zero electron-to-ion mass ratio, two-fluid MHD equations 
simplify and reduce to Hall-MHD equations. The later include 
the Hall and electron pressure terms in the Ohm's law, in 
addition to the resistivity term present in single-fluid
MHD. Because of its relative simplicity, Hall-MHD description of 
plasma has been extensively used in numerical simulations of magnetic 
reconnection~\citep{hesse_2001,fitzpatrick_2004,huba_2004,murphy_2008}. 
However, to the best of our knowledge, a simple self-consistent 
analytical model of Hall reconnection, similar to the classical 
Sweet-Parker model of resistive reconnection, has not been constructed.
In this paper we consider Hall-MHD equations and present a 
theoretical model of Hall reconnection. Full two-fluid calculations 
for magnetic reconnection will be considered elsewhere. The analytical 
derivations of this paper are similar to the derivations done by 
\citet{malyshkin_2005} for the case of resistive single-fluid MHD 
reconnection.

%---------------------------------------------------------------------

\section{\label{HALL}
Hall-MHD equations
}
For simplicity and brevity, 
we use physical units in which the speed of light $c$ and four times 
$\pi$ are replaced by unity, $c=1$ and $4\pi=1$. To rewrite our 
equations in the CGS units, one needs to make the following 
substitutions: magnetic field 
${\bf B}\rightarrow {\bf B}/\sqrt{4\pi}$, electric field 
${\bf E}\rightarrow c{\bf E}/\sqrt{4\pi}$, electric current 
${\bf j}\rightarrow\sqrt{4\pi}\,{\bf j}/c$, electrical resistivity 
$\eta\rightarrow\eta c^2/4\pi$, and the proton electric charge 
$e\rightarrow\sqrt{4\pi}\,e/c$.

We assume the plasma is non-relativistic, with both phase and physical 
velocities much smaller than the speed of light. We neglect
electron inertia for the Hall-MHD description of plasma.
The generalized Ohm's law is~\citep{sturrock_1994} 
\begin{eqnarray}
{\bf E}&=&-{\bf V}\times{\bf B}+\eta{\bf j}
+\frac{m_i}{\rho e}{\bf j}\times{\bf B}
-\frac{m_i}{\rho e}{\bf\nabla}\cdot P_e,
\label{OHMS_LAW}
\end{eqnarray}
where $m_i$ is the ion mass, $\rho$ is the plasma density, 
${\bf V}$ is the plasma velocity (equal to the ion velocity), 
$P_e$ is the tensor of the electron pressure.
The first two terms on the right-hand side of Eq.~(\ref{OHMS_LAW})
are the single-fluid MHD terms, the third and fourth terms are 
the Hall and electron pressure terms.
The equation of plasma motion is~\citep{sturrock_1994}
\begin{eqnarray}
\rho\frac{\partial{\bf V}}{\partial t}
+\rho({\bf V}\cdot{\bf\nabla}){\bf V}=-{\bf\nabla}\cdot P
+{\bf j}\times{\bf B},
\label{MOTION_LAW}
\end{eqnarray}
where $P$ is the tensor of the total pressure (equal to the 
sum of the electron and ion pressure tensors), and we neglect
plasma viscosity. Equation~(\ref{MOTION_LAW}) appears
exactly the same as in the case of single-fluid MHD. Note 
that ${\bf\nabla}\cdot{\bf B}=0$, and, for 
non-relativistic plasma, ${\bf\nabla}\cdot{\bf j}=0$.

\begin{figure}[t]
\vspace{3.6truecm}
\includegraphics{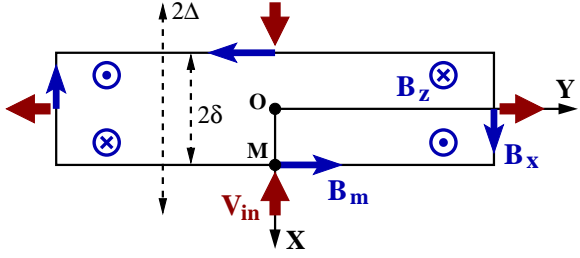}
\caption{Geometrical configuration of the reconnection layer.
}
\label{FIGURE_LAYER}
\end{figure}

We consider Hall magnetic reconnection in the classical 
two-dimensional Sweet-Parker-Petschek reconnection layer, shown in
Fig.~\ref{FIGURE_LAYER}. The layer is in the x-y plane with the
x- and y-axes being perpendicular to and along the layer
respectively. All $\partial/\partial z$ derivatives are zero.
The thickness of the reconnection layer is $2\delta$, 
which is defined as the thickness of the out-of-plane current ($j_z$) 
profile across the layer. Note that  $2\delta$ is approximately equal 
to the electron layer thickness, while the ion layer thickness $2\Delta$ 
can be much larger.
Velocity $V_{in}$ is the plasma inflow velocity in the upstream 
region at point~$M$, outside the electron layer. 
The magnetic field $B_m$ at point~$M$ is in the y-direction. 
The out-of-plane field $B_z$ is assumed to have a quadrupole structure
(see Fig.~\ref{FIGURE_LAYER}), in agreement with 
numerical simulations and laboratory experiments of two-fluid 
reconnection~\citep{mandt_1994,shay_1999,hesse_2001,bhattacharjee_2001,
daughton_2004,fitzpatrick_2004,horiuchi_2004,huba_2004,shay_2004,
yamada_2006,murphy_2008} (a nearly uniform ``guide'' 
field component of $B_z$ is taken to be zero). 
The reconnection layer is assumed to have a
point symmetry with respect to its geometric center point~$O$ in
Fig.~\ref{FIGURE_LAYER} and reflection symmetries with
respect to the axes $x$ and $y$. 
Thus, the x-, y- and z-components of ${\bf V}$, ${\bf B}$ and 
${\bf j}$ have the following simple symmetries:
$V_x(\pm x,\mp y)=\pm V_x(x,y)$, $V_y(\pm x,\mp y)=\mp V_y(x,y)$,
$V_z(\pm x,\mp y)=V_z(x,y)$,
$B_x(\pm x,\mp y)=\mp B_x(x,y)$, $B_y(\pm x,\mp y)=\pm B_y(x,y)$,
$B_z(\pm x,\mp y)=-B_z(x,y)$,
$j_x(\pm x,\mp y)=\pm j_x(x,y)$, $j_y(\pm x,\mp y)=\mp j_y(x,y)$ and
$j_z(\pm x,\mp y)=j_z(x,y)$. In the derivations presented below we
will extensively use these symmetries.

%---------------------------------------------------------------------

\section{\label{SOLUTION}
Solution for Hall reconnection
}

Now let us list four assumptions that we make in this study. 
First, we assume that the plasma flow is incompressible inside 
the reconnection layer, $\rho={\rm constant}$. 
Second, we assume that the electrical resistivity $\eta$ is 
constant and very small, so that the Lundquist number is very large. 
Third, we assume that the reconnection process is slow and 
quasi-stationary, so that we can neglect all time-derivatives 
$\partial/\partial t$ in Eqs.~(\ref{OHMS_LAW})--(\ref{MOTION_LAW})
and below. This assumption is satisfied if the reconnection rate is 
slow, $E_z\ll V_AB_m$, and there are no plasma instabilities in the 
reconnection layer. 
Fourth, we assume that the electron and ion pressure tensors are 
isotropic, so that the pressure tensors in 
Eqs.~(\ref{OHMS_LAW}) and~(\ref{MOTION_LAW}) are scalars. 

Using Ampere's law and neglecting the displacement current, 
we find the x- and y-components of the current to be 
$j_x=\partial_y B_z$ and $j_y=-\partial_x B_z$. 
Here and below we use convenient notations 
$\partial_x\equiv\partial/\partial x$, 
$\partial_y\equiv\partial/\partial y$, 
$\partial_{yy}^2\equiv\partial^2/\partial y^2$,
$\partial_{xyy}^3\equiv\partial^3/\partial x\partial y^2$, and etc.
The z-component of the current at the reconnection layer 
central point~$O$ is
\begin{eqnarray}
j_o\equiv (j_z)_o=
\left(\partial_x B_y-\partial_y B_x\right)_o
\approx B_m/\delta,
\label{AMPERES_LAW}
\end{eqnarray}
where we use estimates $(\partial_y B_x)_o\ll (\partial_x B_y)_o$ 
and $(\partial_x B_y)_o\approx B_m/\delta$ at point~$O$.

Next, neglecting the time derivative in Eq.~(\ref{MOTION_LAW}),
the equation for acceleration of plasma along the reconnection 
layer, in the y-direction, is
$\rho({\bf V}\cdot{\bf\nabla})V_y=-\partial_y P+j_zB_x-j_xB_z$.
We calculate the first order partial derivative $\partial/\partial y$ 
of this equation at point~O and obtain
\begin{eqnarray}
\rho{(\partial_y V_y)_o}^2 &=& 
-(\partial_{yy}^2P)_o+j_o(\partial_y B_x)_o
\nonumber\\
&\approx& -B_m(\partial_{yy}^2 B_y)_m+j_o(\partial_y B_x)_o,
\nonumber\\
&=& 2B_m^2/L^2+j_o(\partial_y B_x)_o.
\label{ACCELERATION}
\end{eqnarray}
Here, we used the fact that  
the pressure term is $(\partial_{yy}^2 P)_o\approx
(\partial_{yy}^2B_y^2/2)_m-(\partial_{yy}^2B_z^2/2)_o=
B_m(\partial_{yy}^2 B_y)_m<0$. Thus, the drop of pressure 
$P$ along the layer is equal to the magnetic pressure drop of the parallel
field component outside the layer. This result follows from the force balance
condition for the plasma across the reconnection layer (in analogy with the
Sweet-Parker derivations for slowly inflowing plasma), and its rigorous 
proof can be found in~\citep{malyshkin_2005}. The
last expression in Eq.~(\ref{ACCELERATION}) is obtained by defining the
magnetic field external scale as
\begin{eqnarray}
L^2\equiv -2B_m\left/(\partial_{yy}^2 B_y)_m\right..
\label{L}
\end{eqnarray}
This is the scale of magnetic field just outside the reconnection electron 
layer (at point M) and can be interpreted as the length of the layer.
%Equation~(\ref{ACCELERATION}) demonstrates that plasma is accelerated 
%along the reconnection layer by the pressure force $2B_m^2/L^2$
%and by the magnetic tension force $j_o(\partial_y B_x)_o$.
The $(\partial_y V_y)_o$ derivative on the left-hand-side
of Eq.~(\ref{ACCELERATION}) can be estimated from plasma 
incompressibility condition at the O-point,
\begin{eqnarray}
(\partial_y V_y)_o=-(\partial_x V_x)_o\approx V_{in}/\delta,
\label{V_DERIVATIVE}
\end{eqnarray}
where we use an estimate $(\partial_x V_x)_o\approx-V_{in}/\delta$.

Next, the Faraday's law 
${{\bf\nabla}\times{\bf E}}=-\partial{\bf B}/\partial t$ for
the x- and y-components of a quasi-stationary magnetic field in two 
dimensions is 
$\partial E_z/\partial y=-\partial B_x/\partial t\approx0$
and $\partial E_z/\partial x=\partial B_y/\partial t\approx0$. 
Therefore, the electric field z-component $E_z$ is constant
in space, and from generalized Ohm's law~(\ref{OHMS_LAW}) 
we obtain
\begin{eqnarray}
\mbox{constant in space} \!&\approx&\! E_z
=-V_xB_y+V_yB_x+\eta j_z
\nonumber\\
&&+(m_p/\rho e)(j_xB_y-j_yB_x).
\qquad
\label{OHMS_LAW_Z}
\end{eqnarray}
Now, we use Eq.~(\ref{OHMS_LAW_Z}) to calculate $E_z$ at points~O 
and~M (see Fig.~\ref{FIGURE_LAYER}). At point~O we have 
$E_z=\eta j_o$. At point~M we have 
$E_z=V_{in}B_m+\eta(j_z)_m+(m_p/\rho e)B_m(j_x)_m\approx
V_{in}B_m+\eta (j_z)_m+(m_p/\rho e)B_m(\partial_{xy}^2B_z)_o\delta$, 
where we use an estimate 
$(j_x)_m\approx(\partial_xj_x)_o\delta=(\partial_{xy}^2B_z)_o\delta$.
Equating these two expressions for $E_z$ at points~O and~M and neglecting
the resistive term outside the reconnection layer at point~M, 
$\eta(j_z)_m\approx\eta B_m/L\ll\eta j_o$, we obtain
\begin{eqnarray}
\eta j_o\approx V_{in}B_m+(m_p/\rho e)B_m(\partial_{xy}^2B_z)_o\delta.
\label{E_Z_PERPENDICULAR}
\end{eqnarray}
Next, we calculate the second order partial derivative 
$\partial^2/\partial y^2$ of Eq.~(\ref{OHMS_LAW_Z}) at point~O. We have
\begin{eqnarray}
0 &\approx& 
2(\partial_y V_y)_o(\partial_y B_x)_o+\eta(\partial_{yy}^2 j_z)_o
\nonumber\\
&& -2(m_p/\rho e)(\partial_y j_y)_o(\partial_y B_x)_o
\nonumber\\
&\approx& 
2(\partial_y V_y)_o(\partial_y B_x)_o-\eta(2j_o/L^2)
\nonumber\\
&& +2(m_p/\rho e)(\partial_{xy}^2B_z)_o(\partial_y B_x)_o.
\label{E_Z_PARALLEL}
\end{eqnarray}
Here, to obtain the final expression, we use the fact that 
the y-scale of the current $j_z$, to a factor of order unity, is 
about the same as the y-scale of the outside magnetic field,
$j_o^{-1}(\partial_{yy}^2j_z)_o
\approx B_m^{-1}(\partial_{yy}^2B_y)_m=-2/L^2$.
This result can be understood by taking the $\partial^2/\partial y^2$ 
partial derivative of the Ampere's law equation $j_o\approx B_m/\delta$, 
see Eq.~(\ref{AMPERES_LAW}), while keeping $\delta$ constant because 
the partial derivative in $y$ is to be taken at constant 
$x=\delta$ (for details see~\citep{malyshkin_2005}).

Next, we use the z-component of the Faraday's law. We have
$0\approx-\partial B_z/\partial t=\partial_x E_y-\partial_y E_x$.
We calculate the $\partial^2/\partial x\partial y$ partial derivative 
of this equation at point~O and use Ohm's law~(\ref{OHMS_LAW}) for 
$E_x$ and $E_y$. After tedious but straightforward derivations, we obtain
\begin{eqnarray}
0 &\approx& 
-\eta\left[(\partial_{xyxx}^4 B_z)_o+(\partial_{xyyy}^4 B_z)_o\right]
\nonumber\\
&&-\left[(\partial_y B_x)_o(\partial_{xx}^2V_z)_o+
(\partial_x B_y)_o(\partial_{yy}^2V_z)_o\right]
\nonumber\\
&&+(m_p/\rho e)\left[(\partial_y B_x)_o(\partial_{xx}^2j_z)_o+
(\partial_x B_y)_o(\partial_{yy}^2j_z)_o\right]
\nonumber\\
&\approx& 
2\eta(\partial_{xy}^2 B_z)_o/\delta^2
\nonumber\\
&&-2(m_p/\rho e)\left[(j_o/\delta^2)(\partial_y B_x)_o+
j_o^2/L^2\right].
\label{E_XY}
\end{eqnarray}
Here, to derive the final expression, we use 
$(\partial_{xyxx}^4 B_z)_o\approx-2(\partial_{xy}^2 B_z)_o/\delta^2$,
$(\partial_{xyyy}^4 B_z)_o\ll(\partial_{xyxx}^4 B_z)_o$,
$(\partial_{xx}^2j_z)_o\approx-2j_o/\delta^2$,
$(\partial_{yy}^2j_z)_o\approx-2j_o/L^2$ and $(\partial_x B_y)_o\approx j_o$.
We also use formula $(\partial_y B_x)_o(\partial_{xx}^2V_z)_o+
(\partial_x B_y)_o(\partial_{yy}^2V_z)_o\equiv0$. To prove it, let us
consider the z-component of plasma motion equation~(\ref{MOTION_LAW}),
which is $\rho V_x(\partial_x V_z)+\rho V_y(\partial_y V_z)=j_xB_y-j_yB_x$.
We calculate the $\partial^2/\partial x^2$ and $\partial^2/\partial y^2$ 
derivatives of this equation at point~O and obtain
$2\rho(\partial_x V_x)_o(\partial_{xx}^2V_z)_o=
2(\partial_x j_x)_o(\partial_x B_y)_o$ and
$2\rho(\partial_y V_y)_o(\partial_{yy}^2V_z)_o=
-2(\partial_y j_y)_o(\partial_y B_x)_o$. Thus, we have
$(\partial_{xx}^2V_z)_o
%=(\partial_x j_x)_o(\partial_x B_y)_o/\rho(\partial_x V_x)_o
=-(\partial_{xy}^2 B_z)_o(\partial_x B_y)_o/\rho(\partial_y V_y)_o<0$,
$(\partial_{yy}^2V_z)_o
%=-2(\partial_y j_y)_o(\partial_y B_x)_o/\rho(\partial_y V_y)_o
=(\partial_{xy}^2 B_z)_o(\partial_y B_x)_o/\rho(\partial_y V_y)_o>0$ and
$(\partial_y B_x)_o(\partial_{xx}^2V_z)_o+
(\partial_x B_y)_o(\partial_{yy}^2V_z)_o\equiv0$.

Now, we have six 
equations~(\ref{AMPERES_LAW}),~(\ref{ACCELERATION}),~(\ref{V_DERIVATIVE}),
~(\ref{E_Z_PERPENDICULAR}),~(\ref{E_Z_PARALLEL}) 
and~(\ref{E_XY}), and we have six unknowns: $j_o$, $\delta$,
$V_{in}$, $(\partial_y V_y)_o$, $(\partial_y B_x)_o$ and 
$(\partial_{xy}^2 B_z)_o$. Thus, we can solve for all unknown quantities. 
For convenience of presentation, we express the solution in terms of 
the Alfven velocity $V_A=B_m/\sqrt{\rho}$, the ion inertial length 
$d_i=m_i/e\sqrt{\rho}$ and the Lundquist number $S=LV_A/\eta$. 
The solution is
\begin{eqnarray}
j_o &\approx& \frac{B_m}{L}\left(S\sqrt{3}+\left.2S^2d_i^2\right/L^2\right)^{1/2},
\label{J_O}
\\
\delta &\approx& L\left(S\sqrt{3}+\left.2S^2d_i^2\right/L^2\right)^{-1/2},
\label{DELTA_O}
\\
V_{in} &\approx& \sqrt{3}\,V_A\left(S\sqrt{3}+\left.2S^2d_i^2\right/L^2\right)^{-1/2},
\label{V_in_O}
\\
(\partial_y V_y)_o &\approx& \sqrt{3}\,V_A\big/L,
\label{Vy_Y}
\\
(\partial_y B_x)_o &\approx& 
\frac{B_m}{L}\left(S\sqrt{3}+\left.2S^2d_i^2\right/L^2\right)^{-1/2},
\label{Bx_Y}
\\
(\partial_{xy} B_z)_o &\approx& 2Sd_iB_m\big/L^3,
\label{Bz_XY}
\\
E_z=\eta j_o &\approx& 
V_AB_m\left(S^{-1}\sqrt{3}+\left.2d_i^2\right/L^2\right)^{1/2}.
\label{Ez}
\end{eqnarray}
The last equation gives the reconnection rate $E_z$. In the limit $S\gg1$ 
and $d_i\ll L$ the reconnection rate is slow, $E_z\ll V_AB_m$, and our 
assumption of a quasi-stationary reconnection process is self-consistent. 
Equation~(\ref{Vy_Y}) implies that the ions are accelerated up to 
approximately Alfven velocity $V_A$ along the reconnection layer of 
length $L$. At the same time, the rate of electron acceleration along 
the layer at point~O is 
$(\partial_y V_y^{e})_o=(\partial_y V_y)_o-(m_i/\rho e)(\partial_y j_y)_o
=(\partial_y V_y)_o+(V_Ad_i/B_m)(\partial_{xy} B_z)_o
\approx (V_A/L)(\sqrt{3}+2Sd_i^2/L^2)$, where ${\bf V}^{e}$
denotes the electron velocity.

The ion layer thickness $2\Delta$ can be estimated as follows. In the 
upstream region outside the ion layer at $x=\Delta$ ideal single-fluid 
MHD applies. Therefore, at $x=\Delta$ and $y=0$ the resistive and Hall 
terms in Eq.~(\ref{OHMS_LAW_Z}) can be neglected and $E_z\approx V_RB_m$, 
where $V_R$ is the plasma inflow velocity outside the ion layer. Velocity 
$V_R\approx E_z/B_m$ is called the reconnection velocity. 
It can also be estimated as $V_R\approx (\partial_y V_y)_o\Delta$. Thus, 
\begin{eqnarray}
\Delta \approx\frac{E_z}{(\partial_y V_y)_oB_m} \approx 
\frac{L}{\sqrt{3}}\left(S^{-1}\sqrt{3}+\left.2d_i^2\right/L^2\right)^{1/2}.
\label{DELTA_ION}
\end{eqnarray}

We discuss the solution~(\ref{J_O})-(\ref{DELTA_ION}) in the next section. 
Now, let us make an important remark. Our analytical derivations 
involve an approximate solution of the Hall-MHD equations 
in the infinitesimal neighborhood of line~OM across the reconnection 
electron layer (see Fig.~\ref{FIGURE_LAYER}). All physical quantities, 
$j_o$, $\delta$, $V_{in}$, $(\partial_y V_y)_o$, 
$(\partial_y B_x)_o$, $(\partial_{xy} B_z)_o$, $B_m$ and $L$, are 
defined either at point~O (the layer center) or at 
point~M (the upstream region). In other words, all our derivations 
involve only the upstream region and the layer center, and we do not 
need to consider the downstream region for estimation of the 
reconnection rate. This ``local'' equations approach was first 
developed in~\citep{malyshkin_2005} for single-fluid MHD reconnection 
with anomalous electrical resistivity, and this approach works for 
Hall-MHD reconnection as well. Note however, that we define the 
field scale $L$ by Eq.~(\ref{L}), and its exact value, as well as 
the value of field $B_m$ in the upstream region, depend on the 
``global'' solution of the Hall-MHD equations outside the 
reconnection layer. Both $L$ and $B_m$ enter our model as parameters. 
Determination of their values requires numerical simulations of 
the global field configuration and is not considered here.

%---------------------------------------------------------------------

\section{\label{DISCUSSION}
Discussion
}

When $d_i\ll L/\sqrt{S}=\delta_{\rm SP}$ ($\delta_{\rm SP}$ is 
the Sweet-Parker layer thickness), the 
solution~(\ref{J_O})-(\ref{DELTA_ION}) 
reduces to the Sweet-Parker solution: 
$j_o\approx\sqrt{S}\,B_m/L$, 
$\delta\approx \Delta\approx L/\sqrt{S}=\delta_{\rm SP}$,
$V_{in}\approx V_A/\sqrt{S}$, 
$(\partial_y V_y)_o\approx V_A/L$,
$(\partial_y B_x)_o\approx B_m/L\sqrt{S}$, 
$E_z\approx V_AB_m/\sqrt{S}$ and 
$(\partial_y V_y^{e})_o\approx(\partial_y V_y)_o\approx V_A/L$.

In the opposite limit, when 
$d_i\gg L/\sqrt{S}=\delta_{\rm SP}$ and reconnection 
is in a collisionless Hall regime, we have
\begin{eqnarray}
j_o &\approx& Sd_iB_m/L^2\gg \sqrt{S}\,B_m/L,
\label{J_O_COLLISIONLESS}
\\
\delta &\approx& L^2/Sd_i\ll L/\sqrt{S}=\delta_{\rm SP},
\label{DELTA_O_COLLISIONLESS}
\\
\Delta &\approx& d_i\gg L/\sqrt{S}=\delta_{\rm SP},
\label{DELTA_ION_COLLISIONLESS}
\\
V_{in} &\approx& V_AL/Sd_i\ll V_A/\sqrt{S},
\label{V_in_O_COLLISIONLESS}
\\
(\partial_y V_y)_o &\approx& V_A/L,
\label{Vy_Y_COLLISIONLESS}
\\
(\partial_y B_x)_o &\approx& B_m/Sd_i,
\label{Bx_Y_COLLISIONLESS}
\\
(\partial_{xy} B_z)_o &\approx& 2Sd_iB_m/L^3,
\label{Bz_XY_COLLISIONLESS}
\\
E_z &\approx& (d_i/L)\,V_AB_m\gg V_AB_m/\sqrt{S}.
\label{Ez_COLLISIONLESS}
\end{eqnarray}
The electron acceleration rate along the reconnection layer 
at O-point is 
$(\partial_y V_y^{e})_o\approx Sd_i^2V_A/L^3\gg (\partial_y V_y)_o$.
Fast electron outflow along the layer creates the quadrupole field 
$B_z$~\citep{uzdensky_2006}.
From Eq.~(\ref{Ez_COLLISIONLESS}) we find that the rate of 
collisionless Hall reconnection, $E_z=(d_i/L)\,V_AB_m$, is 
independent of the electrical resistivity $\eta$~\citep{cowley_1985}. 
The reconnection velocity is $V_R\approx E_z/B_m\approx(d_i/L)V_A$.
From Eq.~(\ref{DELTA_ION_COLLISIONLESS}) we find $\Delta\approx d_i$
for the ion layer thickness, which is in agreement with 
experiment~\citep{yamada_2006}.

In it noteworthy that in the absence of collisions, when $\eta\to0$, 
it is an anisotropic electron pressure that balances $E_z$ field at point~O. 
In this case electrons are accelerated by $E_z$ field during time 
$\sim L/V_{eT}$, while they are unmagnetized and are traveling with 
thermal speed $V_{eT}$ inside the electron layer of length $L$ 
\citep{yamada_2009,kulsrud_etal_2005}. 
As a result, the effective resistivity becomes 
$\eta_{\rm eff}\approx d_e^2V_{eT}/L$, where 
$d_e=\sqrt{m_im_e}/e\sqrt{\rho}$ is the electron inertial length 
($m_e$ is the electron mass). Thus, pressure anisotropy and electron 
inertia become important (they will be considered elsewhere).

In the end of this paper, let us briefly discuss possible mechanisms 
of fast reconnection, which is independent of the macroscopic size 
of the reconnecting system. First, note that the collisionless Hall 
reconnection rate $E_z=(d_i/L)\,V_AB_m$ is high when the 
reconnection electron layer length $L$ is microscopically small and 
comparable to $d_i$. Although the value of $L$ cannot be determined 
in our model, numerical simulations find that $L$ can indeed
be much smaller than the macroscopic system 
size~\citep{mandt_1994,shay_1999,bhattacharjee_2001,fitzpatrick_2004,
huba_2004,shay_2004}. For example, $L$ can be as small as 
$\approx10d_i$ during quasi-stationary collisionless reconnection, 
resulting in a very fast reconnection rate 
$E_z\approx0.1\,V_AB_m$~\citep{shay_1999,huba_2004,shay_2004}.

Another possible mechanism of fast reconnection can result from 
the dependence of the reconnection rate on the density $\rho$ and 
temperature $T$ of the plasma. We have 
$\eta\propto T^{-3/2}$~\citep{sturrock_1994}, 
$V_A\propto\rho^{-1/2}$, $S\propto\rho^{-1/2}T^{3/2}$, 
$d_i^2\propto\rho^{-1}$. Therefore, 
$d_i/\delta_{\rm SP}\propto(T/\rho)^{3/4}$, and the 
collisionless Hall reconnection rate is $E_z\propto\rho^{-1}$. 
Thus, even if initially $L\gg d_i$, as reconnection proceeds, 
the plasma is heated up by Joule heating, the plasma temperature rises, 
the plasma density drops, and the reconnection becomes more and more
collisionless and faster and faster. This run-away reconnection 
process can possibly operate in solar corona and Earth's magnetosphere. 
The dependence $E_z\propto\rho^{-1}$ also supports a model of 
self-regulation heating in solar corona~\citep{uzdensky_2007}.

I am especially grateful to Russell Kulsrud for many very helpful discussions
and for pointing out to me the importance of electron pressure anisotropy.
I would also like to thank Ellen Zweibel, Fausto Cattaneo, Masaaki Yamada,
Boon Chye (BC) Low, Eugene Parker, Alex Obabko, Dmitri Uzdensky, Hantao Ji, 
Viacheslav Titov, Amitava Bhattacharjee and the anonymous referee for their 
interest in this work, discussions and for useful comments. This work was 
supported by the NSF Center for Magnetic Self-Organization (CMSO) in 
Laboratory and Astrophysical Plasmas at the University of Chicago.
\\

{\it A note added after publication:} After the publication of his
paper, the author has became aware that Simakov and Chacon (2008) have
independently obtained a Hall reconnection rate formula, which is the
same as Eq.~(17) up to numerical factors of order
unity~\citep{symakov_2008}. Although the final result is basically the
same, the two derivations are significantly different. The author has
used a rigorous local analysis, in which the Hall-MHD equations are
solved only in the upstream region and in the center of the
reconnection layer. By contrast, Simakov and Chacon (2008) have used a
different approach that also considers the downstream region. In
particular, they have assumed that the thickness of the ion layer is
given by $\max\{d_i,\delta\}$, and obtained their reconnection rate
formula as an approximate interpolation between the rates in the
collisional and collisionless regimes. There are also minor
differences in the problem formulation; that of Simakov and Chacon
(2008) includes electron viscosity, which is neglected by the author,
and neglects electron pressure, which is included by the author. 
Expression $(d_i/L)\,V_A$ for the reconnection velocity in the 
collisionless regime was obtained earlier by Stanley W. H. Cowley
(1985)~\citep{cowley_1985}.

%---------------------------------------------------------------------

%---------------------------------------------------------------------

\end{document}